\pdfoutput=1
\documentclass[11pt,twoside,a4paper,cmspaper,final,collab]{cms-tdr}
\def\svnVersion{9146631-D}\def\svnDate{2021/02/21}\def\cmsCernNoTag{CERN-EP-2020-155}\def\cmsCernDate{\today}\def\cmsMessage{"Published in Physics Letters B as \href{http://dx.doi.org/10.1016/j.physletb.2021.136253}{\doi{10.1016/j.physletb.2021.136253}.}"}
\begin{document}\cmsNoteHeader{HIN-19-008}

\newcommand {\PbPb}  {\ensuremath{\text{PbPb}}\xspace}
\newcommand {\minv}  {\ensuremath{m_{\text{inv}}}\xspace}
\newcommand {\vn}  {\ensuremath{v_{\mathrm{n}}}\xspace}
\newcommand {\vTwo}  {\ensuremath{v_{2}}\xspace}
\newcommand {\vThree}  {\ensuremath{v_{3}}\xspace}
\newcommand {\dcaDz}  {\ensuremath{\text{DCA}}\xspace}
\providecommand {\NA}  {\ensuremath{\text{---}}}
\newlength\cmsTabSkip\setlength{\cmsTabSkip}{1ex}
\newlength\cmsFigWidth\ifthenelse{\boolean{cms@external}}{\setlength\cmsFigWidth{0.49\textwidth}}{\setlength\cmsFigWidth{0.65\textwidth}} 

\cmsNoteHeader{HIN-19-008}
\title{Measurement of prompt \texorpdfstring{\PDz}{D0} and \texorpdfstring{\PADz}{anti-D0} meson azimuthal anisotropy and search for strong electric fields in PbPb collisions at \texorpdfstring{$\sqrtsNN=5.02\TeV$}{sqrt(sNN) = 5.02 TeV}}

\date{\today}

\abstract{
The strong Coulomb field created in ultrarelativistic heavy ion collisions is expected to produce a rapidity-dependent difference ($\Delta \vTwo$) in the second Fourier coefficient of the azimuthal distribution (elliptic flow, $\vTwo$) between \PDz ($\PAQu\PQc$) and \PADz ($\PQu\PAQc$) mesons. Motivated by the search for evidence of this field, the CMS detector at the LHC is used to perform the first measurement of $\Delta \vTwo$. The rapidity-averaged value is found to be $\langle\Delta \vTwo \rangle = 0.001\pm0.001\stat\pm0.003\syst$ 
in \PbPb collisions at $\sqrtsNN=5.02\TeV$. In addition, the influence of the collision geometry is explored by measuring the \PDz and \PADz mesons $\vTwo$ and triangular flow coefficient ($\vThree$) as functions of rapidity, transverse momentum (\pt), and event centrality (a measure of the overlap of the two Pb nuclei). A clear centrality dependence of prompt \PDz meson $\vTwo$ values is observed, while the $\vThree$ is largely independent of centrality. These trends are consistent with expectations of flow driven by the initial-state geometry. 
}

\hypersetup{%
pdfauthor={CMS Collaboration},%
pdftitle={Search for strong electric fields in PbPb collisions at sqrt(sNN) = 5.02 TeV using azimuthal anisotropy of prompt D0 and anti-D0 mesons},%
pdfsubject={CMS},%
pdfkeywords={CMS, heavy-flavor, charm, electromagnetic fields}}

\maketitle

\section{Introduction}
\label{intro}

The observation of a strongly-coupled quark-gluon plasma (QGP), a state of matter composed of 
deconfined quarks and gluons, was established by experiments 
investigating ultrarelativistic heavy ion collisions at the BNL RHIC~\cite{Arsene:2004fa,Back:2004je,Adams:2005dq,Adcox:2004mh} 
and CERN LHC~\cite{Muller:2012zq,Armesto:2015ioy}. The azimuthal particle correlations constitute an
effective tool to probe the properties of the QGP~\cite{Arsene:2004fa,Back:2004je,Adams:2005dq,Adcox:2004mh,Muller:2012zq,Armesto:2015ioy,Aamodt:2010pa,ATLAS:2011ah,Chatrchyan:2012ta}. These correlations are parameterized by a Fourier expansion~\cite{Ollitrault:1993,Voloshin:1994,Poskanzer:1998yz},
with the magnitude of the Fourier coefficients, $\vn$, providing information about the initial collision
geometry and its fluctuations~\cite{Poskanzer:1998yz}. 
The second- ($\vTwo$) and third- ($\vThree$) order Fourier coefficients are referred to as 
``elliptic'' and ``triangular'' flow harmonics, respectively. Measuring these coefficients for particle 
species with different quark composition provides additional information about this hot and dense medium~\cite{Molnar:2003ff}. 
Because of their large mass, charm and bottom quarks are produced earlier in the collisions
than the light quarks (up and down)~\cite{BraunMunzinger:2007tn,Liu:2012ax}. 
In addition, the charm and bottom quarks have masses many times larger than the typical temperatures in the QGP~\cite{Borsanyi:2010bp}. 
These heavy quarks experience the full evolution of the medium until the hadronization phase. As a consequence, the 
$\vn$ of charmed \PDz ($\PAQu\PQc$) and \PADz ($\PQu\PAQc$) mesons 
(henceforth referred to as \PDz mesons, except where explicitly stated otherwise) 
are expected to receive important contributions from medium energy loss and coalescence effects~\cite{Moore:2004tg,Andronic:2015wma}.

In ultrarelativistic heavy ion collisions, very strong and transient (${\sim}10^{-1}\unit{fm}/c$) 
magnetic and electric fields are expected to be induced by the collision spectators and participants~\cite{Gursoy:2018yai}. 
Such electromagnetic (EM) fields are predicted to produce a difference in the $\vn$ harmonics for positively and negatively 
charged particles~\cite{Gursoy:2018yai}. 
In such a picture, the magnetic field is mainly responsible for splitting the rapidity ($y$)-odd 
directed flow ($v_1$)~\cite{Gursoy:2018yai,Das:2016cwd}. 
The electric field is predicted to induce a charge-dependent splitting in the $\vTwo$ coefficient 
and in the average transverse momentum ($\langle\pt\rangle$) values of the emitted particles~\cite{Gursoy:2018yai}. 
As charm quarks are expected to be created very early in the collision, 
they have a higher probability of interacting with this strong EM field than the light flavor quarks~\cite{Das:2016cwd, Chatterjee:2017ahy}.

In this letter, measurements of the $v_{2}$ and $v_{3}$ coefficients as 
functions of \PDz meson rapidity, \pt, and lead-lead (\PbPb)
collision centrality are presented. The collision centrality bins are given 
in percentage ranges of the total inelastic hadronic cross section, 
with the 0--10\% centrality bin corresponding to the 10\% of collisions having the largest overlap of the two nuclei. 
The flow harmonics are measured using the scalar product method~\cite{Adler:2002pu,Luzum:2012da}. 
In this analysis, the selection of \PDz mesons uses multivariate methods~\cite{Hocker:2007ht}
for selecting \PDz candidates and their antiparticles. 
The contamination from nonprompt \PDz candidates, arising from \PB\ meson decay, is considered as a systematic uncertainty. 
Using the data recorded in \PbPb collisions during the 2018 LHC run period, corresponding to $0.58\nbinv$ of 
integrated luminosity, the flow coefficients are measured within the rapidity range $\abs{y}<2$, 
which is twice as large as achieved in previous CMS measurements~\cite{Sirunyan:2017plt}.
The extension of the measurements to this larger rapidity range, 
together with smaller statistical uncertainties provided by a larger data set, 
furnish important inputs for a better understanding of the three-dimensional evolution of the QGP formed in heavy ion collisions. 
Measurements of the $v_{2}$ difference between \PDz and \PADz mesons, $\Delta v_{2}$, as 
a function of rapidity are presented as a method to probe possible effects originating from the Coulomb fields. 

\section{Experimental apparatus and data sample}
\label{sec:detector}

The central feature of the CMS apparatus is a superconducting solenoid of 6\unit{m}
internal diameter, providing a magnetic field of 3.8\unit{T}. Within the solenoid volume,
there are four primary subdetectors including a silicon pixel and strip tracker detector, 
a lead tungstate crystal electromagnetic calorimeter, and a brass and scintillator hadron 
calorimeter, each composed of a barrel and two endcap sections. Iron and quartz-fiber 
Cherenkov hadron forward (HF) calorimeters cover the pseudorapidity range $2.9<\abs{\eta}<5.2$. 
The HF calorimeters are segmented to form $0.175{\times}0.175$ ($\Delta\eta{\times}\Delta\phi$) towers.
Muons are measured in gas-ionization detectors embedded in the steel flux-return yoke 
outside the solenoid. The silicon tracker measures charged particles within the range 
$\abs{\eta}<2.5$. A detailed description of the 
CMS detector, together with a definition of the coordinate system used and the relevant 
kinematic variables, can be found in Ref.~\cite{Chatrchyan:2008zzk}.

The analysis presented in this letter uses approximately $4.27\ten{9}$ 
minimum bias (MB) \PbPb collision events collected by the CMS experiment during the 2018 LHC run. 
The MB events are triggered by requiring signals in both forward and backward sides of the HF calorimeters~\cite{Khachatryan:2016bia}. 
Further selections are applied offline to reject events from background processes (beam-gas interactions and nonhadronic collisions), 
see Ref.~\cite{Khachatryan:2016odn} for details. 
Events are required to have at least one interaction vertex, reconstructed based on two tracks or more, 
and with a distance of less than 15\cm from the center of the nominal interaction point along the beam axis. 
The primary interaction vertex is defined as the one with the highest track multiplicity in the event.
The shapes of the clusters in the pixel detector have to be
compatible with those expected from particles produced at the primary vertex location.
The \PbPb collision events are also required to have at least two calorimeter towers in each HF detector
with energy deposits of more than 4\GeV per tower. These criteria select $(99\pm2)\%$ of inelastic
hadronic \PbPb collisions. The possibility to have values higher than 100\% reflects the possible presence
of ultra-peripheral (nonhadronic) collisions in the selected event sample.

Events from Monte Carlo (MC) simulations are used to study both prompt and nonprompt \PDz 
meson processes. The events are generated using an embedding procedure, in which \PDz mesons 
generated by \PYTHIA 8.212~\cite{Sjostrand:2014zea} (tune CP5~\cite{Sirunyan:2019dfx}) are embedded into MB events from \HYDJET 
1.9~\cite{Lokhtin:2005px}. A full simulation of the CMS detector is performed using \GEANTfour~\cite{Agostinelli:2002hh}.  
The prompt \PDz meson MC simulation is employed to define signal selections 
and measure efficiency corrections, while the nonprompt \PDz meson MC sample is used to estimate 
systematic uncertainties coming from nonprompt \PDz contamination.  

\section{Reconstruction and selection of \texorpdfstring{\PDz}{D0} mesons}
\label{D0recoSel}

Prompt \PDz mesons are reconstructed from the decay 
$\PDz\to\PGpp+\PKm$ and $\PADz\to\PGpm+\PKp$ with a branching fraction of $(3.94\pm0.04)\%$, 
using selected tracks with $\pt>1.0\GeVc$ and within the acceptance of $\abs{\eta}<2.4$.
Candidates are formed by combining pairs of tracks from oppositely charged particles and requiring 
an invariant mass ($\minv$) within a ${\pm}200\MeVcc$ window of the world-average 
\PDz meson mass of $(1864.83\pm0.05)\MeVcc$~\cite{Tanabashi:2018oca}. 
For each pair of selected tracks, two possible candidates for \PDz and \PADz 
mesons are considered by assuming one of the tracks has the pion mass, while 
the other track has the kaon mass, and vice versa. Kinematic vertex fits are performed 
to reconstruct the secondary vertices of \PDz candidate decays.

After the \PDz candidate reconstruction, a selection using a boosted decision tree (BDT) 
algorithm from the \textsc{tmva} package~\cite{Hocker:2007ht} is employed. 
For the BDT training, misidentified \PDz candidates in data events, where pion and kaon have the same charge, 
are used to mimic the combinatorial background. The signal candidates are taken from MC simulations of prompt \PDz mesons and are required 
to match \PDz particles at the generator level. The variables related to \PDz mesons used to 
discriminate the signal from the background  are:
$\chi^2$ probability for the \PDz vertex fit, 3D distance between the secondary and primary vertices and its significance, 
the decay length significance projected in the $xy$-plane, and the angle in two and three dimensions between 
the momentum of the \PDz meson candidate and the line connecting the primary and the 
secondary vertices (pointing angle). Related to the decay products of the \PDz meson candidate, the variables used are:
the uncertainty in \pt returned by the track fitting procedure, 
the significance of the $z$ and the $xy$ distances of closest approach to the primary vertex, 
and the number of hits in the tracker detector. 
These variables are chosen by analyzing their BDT ranking (variables more frequently used in the decision tree) and 
correlation matrix among all variables.
Different BDT boost algorithms are tested, choosing the adaptive boost algorithm~\cite{Hocker:2007ht} as default.  
Overtraining checks are done for all analysis bins by comparing the BDT distributions from training and testing \PDz meson samples.
In addition, a BDT cut optimization is performed in bins of 
centrality, \pt, and rapidity, doing a scan in different BDT scores and finding the one resulting in maximal
\PDz mesons signal significance for each analysis bin. 
Compared to a cutoff-based procedure, this 
BDT selection almost doubles the signal significance for \PDz mesons in $1<\abs{y}<2$, and 
increases the signal significance by 30\% for \PDz mesons in $\abs{y}<1$, for events with collision 
centrality in the range 0--30\%. For the remaining analysis bins a similar performance of BDT and cutoff-based methods is observed.

\section{Analysis technique}
\label{analysisTech}

The elliptic and triangular flow coefficients of \PDz 
mesons are extracted using the scalar product (SP) method, similarly to what was done in a
previous CMS publication~\cite{Sirunyan:2017plt}.
In this method, the $\vn$ coefficients of \PDz candidates (including backgrounds) are measured using
\begin{linenomath*}
\begin{equation}
\label{eq:scalar_product_vn}
\centering
\vn \{ \mathrm{SP} \} \equiv \frac{ \langle Q_{\mathrm{n}}^{\PDz} Q_{\mathrm{nA}}^{*} \rangle }{\sqrt{\smash[b]{\frac{ \langle Q_{\mathrm{nA}}Q_{\mathrm{nB}}^{*} \rangle \langle Q_{\mathrm{nA}}Q_{\mathrm{nC}}^{*} \rangle }{ \langle Q_{\mathrm{nB}}Q_{\mathrm{nC}}^{*} \rangle }}}},
\end{equation}
\end{linenomath*}
with the $Q$-vectors expressed as $Q_{\mathrm{n}} \equiv \sum_{\mathrm{j}=1}^{\mathrm{M}} w_{\mathrm{j}} e^{in\phi_{\mathrm{j}}}$, 
where the sum is over the total number ($\mathrm{M}$) of HF towers above a certain energy threshold 
(with the weights $w_{\mathrm{j}}$ taken as the energy deposited in the HF tower at azimuthal angle $\phi_{\mathrm{j}}$), 
of tracks with \pt above a certain threshold (with $w_{\mathrm{j}}$ taken as track \pt in $\phi_{\mathrm{j}}$ angle), 
or of selected \PDz meson candidates (with $w_{\mathrm{j}}$ taken equal to 1).

The $Q$-vectors related to HF and the tracker are measured and corrected for 
detector irregularities by applying a flattening and a recentering procedure~\cite{Poskanzer:1998yz,Alt:2003ab}.
The $Q_{\mathrm{nA}}$ and $Q_{\mathrm{nB}}$ are defined using the event-plane measurements 
from the negative ($-5<\eta<-3$, $\mathrm{HF-}$) and the positive ($3<\eta<5$, $\mathrm{HF+}$) sides 
of HF, and $Q_{\mathrm{nC}}$ is measured using the tracker information in the region of 
$\abs{\eta}<0.75$, allowing to minimize the correlations among the three regions, with a gap of more than two units of rapidity. 
The $Q_{\mathrm{n}}^{\PDz}$ vector is defined for each \PDz meson candidate. The averages 
$\langle Q_{\mathrm{nA}}Q_{\mathrm{nB}}^{*} \rangle$, $\langle Q_{\mathrm{nA}}Q_{\mathrm{nC}}^{*} \rangle$, and 
$\langle Q_{\mathrm{nB}}Q_{\mathrm{nC}}^{*} \rangle$ are made considering all selected events, while the average 
$\langle Q_{\mathrm{n}}^{\PDz} Q_{\mathrm{nA}}^{*} \rangle$ is made considering all \PDz meson candidates in all selected events.
To avoid autocorrelations, the terms $\langle Q_{\mathrm{n}}^{\PDz} Q_{\mathrm{nA}}^{*} \rangle$ and 
$\langle Q_{\mathrm{nA}}Q_{\mathrm{nB}}^{*} \rangle$ use $\mathrm{A} = \mathrm{HF-}$ ($\mathrm{HF+}$) when 
the \PDz meson candidate is at positive (negative) pseudorapidity. 

\begin{figure*}[h!t]
  \centering
   \includegraphics[width=0.49\textwidth]{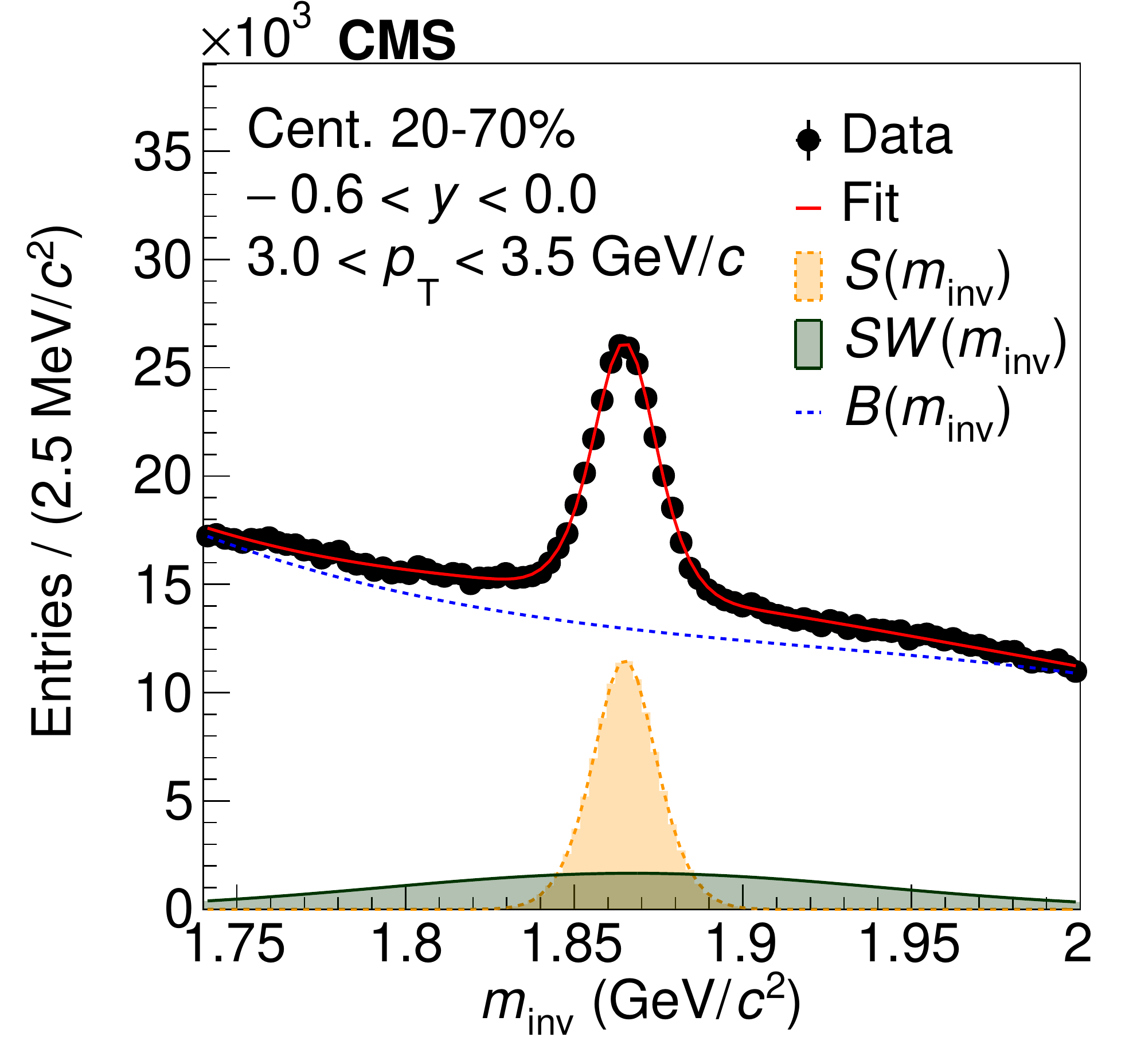}
   \includegraphics[width=0.49\textwidth]{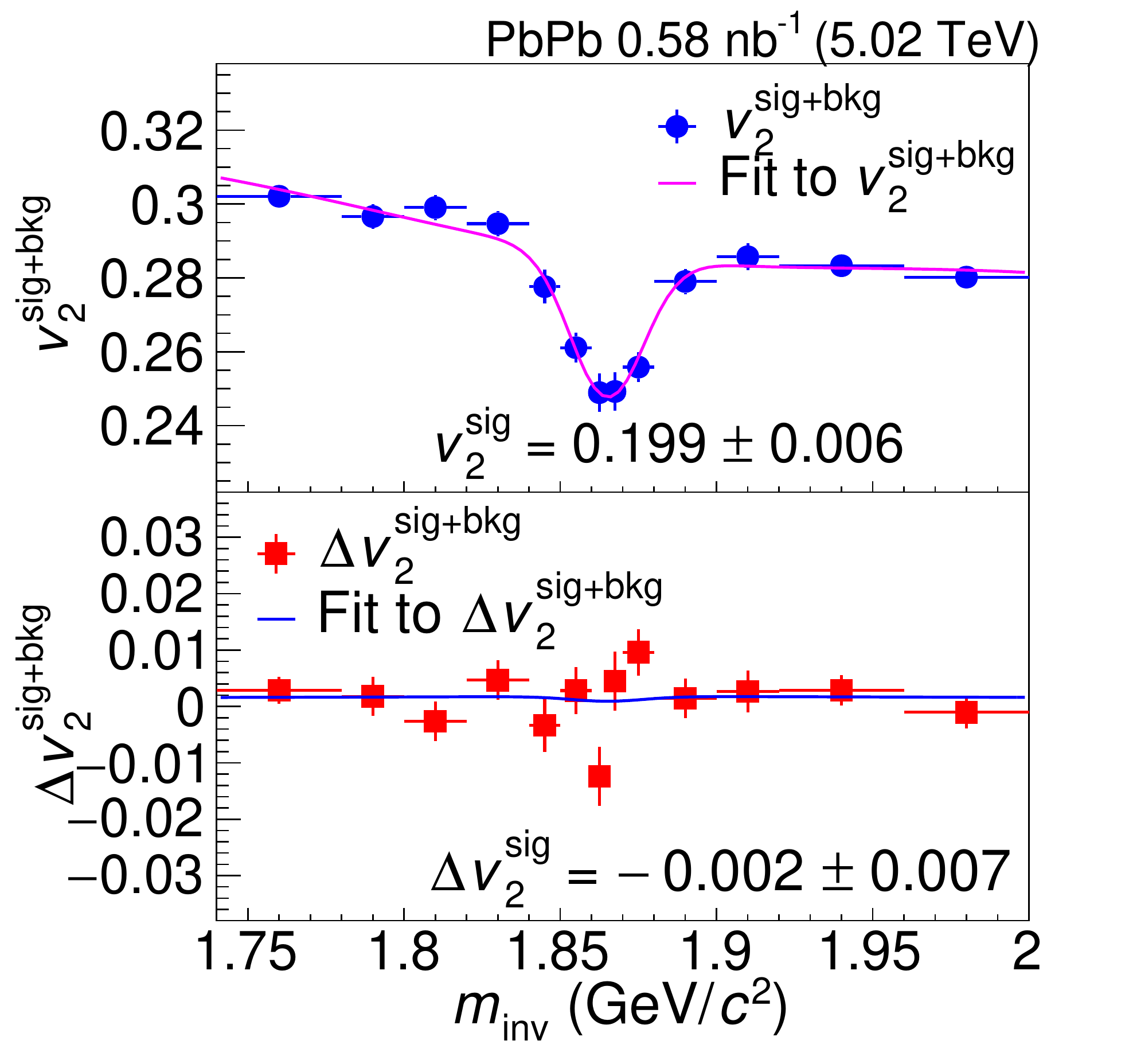}
   \caption{Simultaneous fit of the $\PGp\PK$ invariant mass (left) and $\vTwo$ ($\Delta \vTwo$) as function of invariant mass (right) for $3.0<\pt<3.5\GeVc$, centrality 20--70\%, and $-0.6<y<0.0$.}
   \label{fig:mass_vn_fit}
\end{figure*}

One goal of this analysis is to measure the difference ($\Delta \vn$) between \PDz and \PADz meson flow coefficients, $\vn$, 
as a function of rapidity, to probe effects from EM fields. The difference $\Delta \vn$ 
is measured as:
\begin{linenomath*}
\begin{equation}
\label{eq:scalar_product_dvn}
\centering
\Delta \vn \{ \mathrm{SP} \} \equiv \frac{ \langle Q_{\mathrm{n}}^{\PDz} Q_{\mathrm{nA}}^{*} \rangle - \langle Q_{\mathrm{n}}^{\PADz} Q_{\mathrm{nA}}^{*} \rangle}{\sqrt{\smash[b]{\frac{ \langle Q_{\mathrm{nA}}Q_{\mathrm{nB}}^{*} \rangle \langle Q_{\mathrm{nA}}Q_{\mathrm{nC}}^{*} \rangle }{ \langle Q_{\mathrm{nB}}Q_{\mathrm{nC}}^{*} \rangle }}}}.
\end{equation}
\end{linenomath*}
The $\vn$ and $\Delta \vn$ of \PDz meson candidates are first measured as a function of their $\minv$.
The extraction of the \PDz mesons signal $\vn$ ($\Delta \vn$), $\vn^{\text{sig}}$ ($\Delta \vn^{\text{sig}}$), 
is performed via a simultaneous binned $\chi^2$ fit of the $\minv$ distribution and of $\vn$ ($\Delta \vn$). 
The $\minv$ distribution is fit with three components: a third-order polynomial to model the combinatorial background, $B(\minv)$; 
two Gaussians with the same mean but different widths to describe the $\minv$ in different kinematic regions for the \PDz mesons signal, 
$S(\minv)$; and one additional Gaussian distribution for the swap component corresponding to the 
incorrect mass assignment for the assumed pion and kaon particles, $SW(\minv)$. 
The width of $SW(\minv)$ and the ratio between the yields of $SW(\minv)$ and $S(\minv)$ are 
fixed by the values extracted from MC simulations.
In this case, the following expression can be used for extracting $\vn^{\text{sig}}$:
\begin{linenomath*}
\begin{equation}
\label{eq:vn_signal}
\vn^{\text{sig} + \text{bkg}}(\minv) = \alpha(\minv)\vn^{\text{sig}} + [1 - \alpha(\minv)]\vn^{\text{bkg}}(\minv). 
\end{equation}
\end{linenomath*}
\begin{linenomath*}
The $\alpha(\minv)$ parameter, which characterizes the signal fraction as a function of mass, is defined as follows:
\ifthenelse{\boolean{cms@external}}
{
\begin{multline}
\label{eq:alpha}
\alpha(\minv) = [ S(\minv) + SW(\minv) ]/[S(\minv) + SW(\minv)\\ 
+ B(\minv) ] = \alpha^{\text{signal}}(\minv) + \alpha^{\text{swap}}(\minv). 
\end{multline}
}
{
\begin{equation}
\label{eq:alpha}
\begin{split}
\alpha(\minv) & = [ S(\minv) + SW(\minv) ]/[ S(\minv) + SW(\minv) + B(\minv) ] \\
                & = \alpha^{\text{signal}}(\minv) + \alpha^{\text{swap}}(\minv). 
\end{split}
\end{equation}
}
\end{linenomath*}
For extracting the difference $\Delta \vn^{\text{sig}}$, the following expression is employed:
\begin{linenomath*}
\begin{equation}
\label{eq:dvn_signal}
\Delta \vn^{\text{sig} + \text{bkg}}(\minv) = \Delta \vn^{\text{sig}}(\alpha^{\text{signal}}(\minv) - \alpha^{\text{swap}}(\minv)) + const. 
\end{equation}
\end{linenomath*}
The term $\vn^{\text{bkg}}(\minv)$ from Eq.~(\ref{eq:vn_signal}) is modeled with a linear function, 
while the constant parameter $const$ in Eq.~(\ref{eq:dvn_signal}) is added to 
account for possible fluctuations in the background $\vn$ component.
The relevance of this $const$ parameter was investigated by redoing $\Delta \vn$ measurements in MC simulation 
(without azimuthal correlations or effects from EM fields), 
indicating that this parameter improves the fit quality and 
does not introduce artificial signals. 
A cross-check is performed by redoing the measurements using a linear function 
instead of a constant. No significant changes in the central values of $\Delta \vTwo$ and on their uncertainties are observed.
Figure~\ref{fig:mass_vn_fit} shows an example of a simultaneous fit for $\vTwo$ and $\Delta \vTwo$.

After performing the fits for extracting the signal $\vn$, there is still 
a sizable fraction of nonprompt \PDz mesons embedded in $\vn^{\text{sig}}$. 
The extracted $\vn$ can be written as
\begin{linenomath*}
\begin{equation}
\label{eq:vn_nonprompt}
\vn^{\text{sig}} = f_{\text{prompt}}\vn^{\text{prompt}} + (1 - f_{\text{prompt}})\vn^{\text{nonprompt}}.
\end{equation}
\end{linenomath*}
The nonprompt \PDz meson contamination is taken into account as a systematic uncertainty, 
by checking that the nonprompt \PDz meson fraction is always smaller than 12\% 
(i.e., comparable to the uncertainties in the reconstructed \PDz meson yield). 
This implies that the central values of $\vn$ will not be considerably affected by this component, 
being compatible within statistical uncertainties.
Such a low fraction arises from the use of prompt \PDz meson signals in the BDT training, 
together with variables that are highly correlated with the distance of closest approach ($\dcaDz$) 
to the primary vertex, which is defined as the flight distance of the \PDz particle times the sine of the pointing angle in three dimensions. 
Additional DCA selection and dedicated training, involving prompt and nonprompt \PDz meson signals, 
do not bring considerable improvements in performance. 
The prompt and nonprompt \PDz meson fractions are obtained using the $\dcaDz$ variable.
For prompt \PDz mesons, the nonzero $\dcaDz$ corresponds to the detector resolution, and is expected to be concentrated around zero. For
nonprompt \PDz mesons, larger values of $\dcaDz$ result from the \PB\ meson decay.
To extract the prompt and nonprompt \PDz meson fractions, a fit to the $\dcaDz$ distributions is performed in data 
considering $\dcaDz$ shapes from MC simulations for prompt and nonprompt \PDz meson components.
The nonprompt \PDz meson $\vn$ is estimated by considering two regions in the $\dcaDz$: one with 
very low fraction (2.7--8.0\%) of nonprompt \PDz particles ($\dcaDz<0.012\cm$), and one 
with a high fraction (62.0--88.0\%) of nonprompt \PDz particles ($\dcaDz>0.012\cm$). Using 
this information together with Eq.~(\ref{eq:vn_nonprompt}), it is possible to estimate 
$\vn^{\text{nonprompt}}$ by solving a system of two equations from the two $\dcaDz$ regions. 
In the current analysis this procedure can only be done in wide \pt, centrality, and rapidity bins, 
because of the limited amount of data available in the region with $\dcaDz>0.012\cm$. 

\section{Systematic uncertainties}
\label{systematics}

The sources of systematic  uncertainties include the \PDz identification requirements (BDT selection); the
probability distribution function (PDF) for modeling the background in the 
invariant mass fit; 
the impact of acceptance and efficiency of the \PDz meson yield; 
the variation of the PDF for modeling the background $\vn$; and the remaining nonprompt \PDz contamination.
With the exception of the last component, the uncertainties are quoted as 
 absolute values of $\vn$ and $\Delta \vn$ after comparing the default analysis configuration with the variations. 
To diminish the influence of statistical fluctuations, after observing no special trends 
in the deviations from the default measurements, the systematic uncertainties are evaluated by 
averaging the deviations with a constant fit as a function of the analysis bins.

In order to take into account the systematic uncertainty associated with the BDT selection, 
the BDT cut is varied up and down by the maximal deviation between the BDT optimized selection 
based on MC simulations and data. The BDT cuts (and variations for systematic uncertainties) are defined 
in bins of collision centrality, \pt, and rapidity, ranging from 0.28 to 0.47 (${\pm}$0.02--0.03). 
Regarding the effect of the  background mass modeling, either an exponential function 
together with a second order polynomial, or just a second order polynomial, are considered 
instead of the default fit function using a third-order polynomial. 
To fit $\vn$ as a function of mass, 
the default configuration using a linear function is replaced by 
either a constant or a second order polynomial. 
Although the \PDz meson selection efficiency essentially cancels in $\vn$ measurements, 
a systematic uncertainty is assigned by comparing the results with and without applying 
corrections based on MC simulations in bins of \pt and rapidity. 
The \PDz meson selection efficiency times acceptance varies from 0.5 to 12.5\% 
in the \pt range of 1.0--8.0$\GeVc$, reaching a plateau of approximately 17.0\% for $\pt>15.0\GeVc$.

The systematic uncertainties regarding contamination from nonprompt \PDz 
mesons are estimated by measuring nonprompt \PDz meson $\vn$ in wide 
bins of \pt, rapidity, and centrality. 
A relative systematic uncertainty is obtained by comparing $\vn$ from
mixed prompt and nonprompt \PDz mesons to the $\vn$ derived from nonprompt \PDz mesons. 

Table~\ref{tab:summary_sys_v2_v3_deltav2} summarizes 
the estimates of systematic uncertainties in absolute values for $\vTwo$, $\vThree$, and $\Delta \vTwo$.
The ranges of variation of the uncertainties are presented for each binning.

\begin{table*}[h!t]
 \topcaption{Summary of systematic uncertainties in absolute values for $\vTwo$, $\vThree$, and $\Delta \vTwo$. Ranges of the variation of uncertainties for all the bins are presented. The cells filled with ``\NA'' refer to the cases where the uncertainty cancels out.}
 \label{tab:summary_sys_v2_v3_deltav2}
 \centering 
 \begin{tabular}{lllllll}
  \hline
 Systematic sources                                      & \pt bins & $y$ bins & Centrality bins                     \\
  \hline
                                          & \multicolumn{3}{c}{ $\vTwo$ }  \\[\cmsTabSkip]
  BDT selection                           & 0.002--0.014   & 0.0065   & 0.005   \\
  Bkg. mass PDF                           & 0.0002--0.0017 & 0.0007--0.0015 & 0.0007--0.0011 \\
  Bkg. $\vn$ PDF                          & 0.01--0.05   & 0.004--0.007   & 0.003--0.005      \\
  \PDz efficiency correction             & \NA              & 0.004--0.007   & 0.0040--0.0045         \\
  Nonprompt \PDz meson contamination & 0.0002--0.0077   & 0.004         & 0.002--0.005     \\[\cmsTabSkip]
                                          & \multicolumn{3}{c}{ $\vThree$ }  \\[\cmsTabSkip]
  BDT selection                           & 0.002--0.023   & 0.001--0.009           & 0.002--0.006   \\
  Bkg. mass PDF                           & 0.0001--0.0040 & 0.0005--0.0008 & 0.0012--0.0040 \\
  Bkg. $\vn$ PDF                          & 0.01--0.05  & 0.003--0.004   & 0.0011        \\
  \PDz efficiency correction             & \NA              & 0.002--0.004   & 0.003--0.005            \\
  Nonprompt \PDz meson contamination & 0.0001--0.0090 & 0.0010--0.0015        & 0.0001--0.0008           \\[\cmsTabSkip]
                                          &  \multicolumn{3}{c}{ $\Delta \vTwo$ }                \\[\cmsTabSkip]
  BDT selection                           & & 0.001--0.009 &   \\
  Bkg. mass PDF                           & & 0.00015--0.00030  &  \\
  \PDz efficiency correction             & & 0.001--0.004 &              \\
  Nonprompt \PDz meson contamination & & 0.00002--0.00010 &          \\
  \hline
 \end{tabular}
\end{table*}

\section{Results}
\label{results}

\begin{figure*}[thb]
  \centering
   \includegraphics[width=0.98\textwidth]{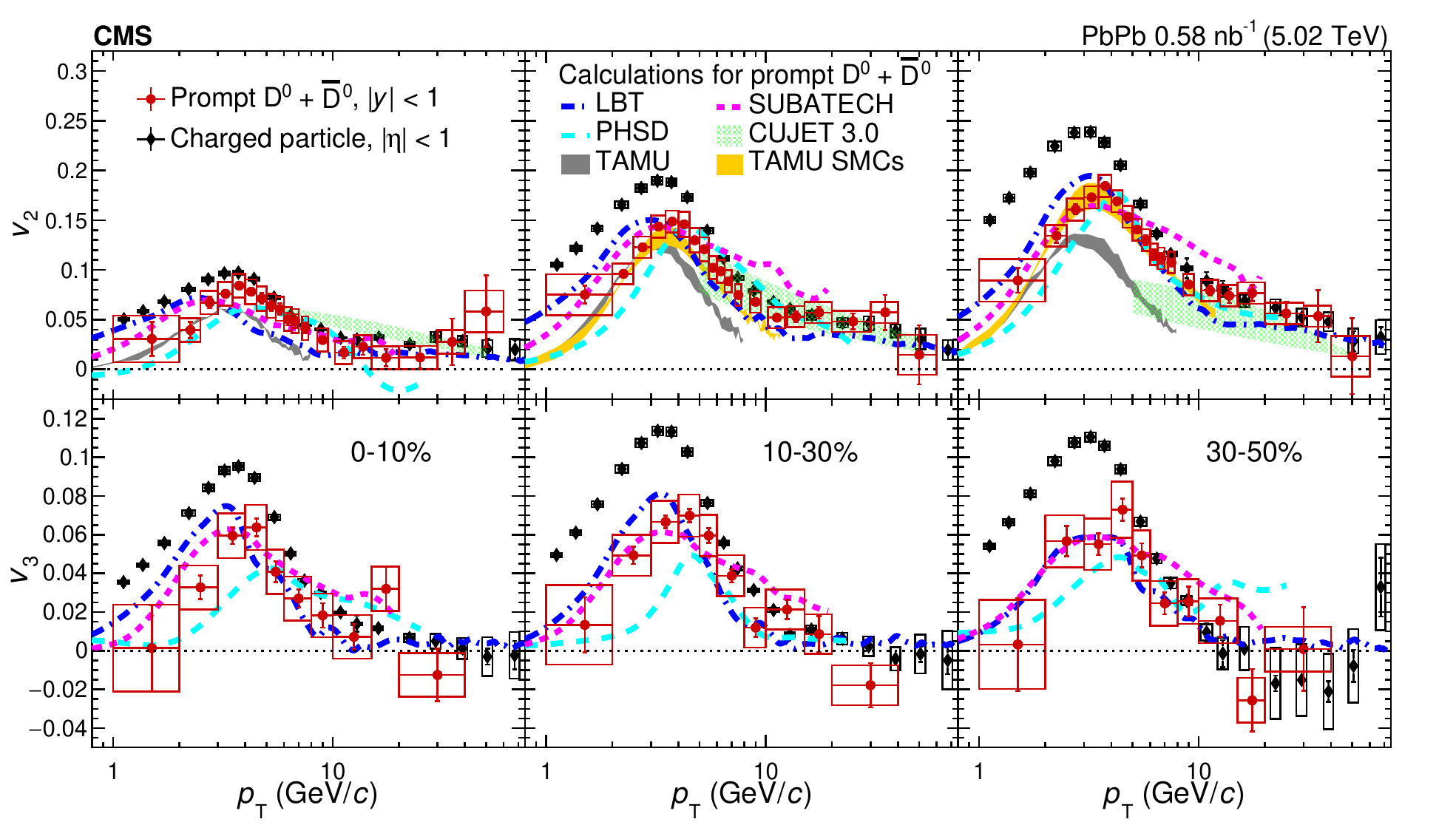}
  \caption{
Prompt \PDz meson and charged particle flow coefficients $\vTwo$ (upper) and $\vThree$ (lower) at midrapidity 
($\abs{y}<1.0$ for prompt \PDz mesons and $\abs{\eta}<1.0$ for charged particles) 
for the centrality classes 0--10\% (left), 10--30\% (middle), and 30--50\% (right). 
The vertical bars and open boxes represent the statistical and systematic uncertainties, respectively.
The horizontal bars represent the width of each \pt bin.
Theoretical calculations for $\vn$ coefficients of prompt \PDz mesons are also plotted for comparison: 
LBT~\cite{Cao:2016gvr}, CUJET 3.0~\cite{Xu:2015bbz}, SUBATECH~\cite{Nahrgang:2014vza}, TAMU~\cite{He:2014cla}, PHSD~\cite{Liu:2012ax}. 
The TAMU SMCs model~\cite{He:2019vgs} is available only in the 10--50\% centrality bins.
  }
  \label{fig:results_vn_pTbins_barrel}
\end{figure*}

Results for prompt \PDz meson $\vTwo$ and $\vThree$ anisotropic flow coefficients, obtained with 2018 \PbPb data, 
as functions of \pt and for $\abs{y}<1$, are shown in Fig.~\ref{fig:results_vn_pTbins_barrel} 
for three centrality ranges: 0--10\%, 10--30\%, and 30--50\%. 
The results extend previously published data from CMS~\cite{Sirunyan:2017plt}, 
by extending the high-\pt coverage to ${\sim}60.0\GeVc$ and by providing finer \pt bins. 
These high-precision data are compatible with previous measurements from Ref.~\cite{Sirunyan:2017plt}, 
and a clear trend of rise and fall from low to high \pt is observed for both $\vTwo$ and $\vThree$ across the 
full centrality range. This behavior is similar to that observed for inclusive charged 
particles~\cite{Sirunyan:2017pan} for $\abs{\eta}<1.0$, also shown in Fig.~\ref{fig:results_vn_pTbins_barrel}. 
For noncentral collisions (i.e., centrality 10--50\%), values of 
prompt \PDz meson $\vTwo$ are positive up to $\pt\sim30.0\text{--}40.0\GeVc$, 
whereas the $\vThree$ values become consistent with zero at $\pt\sim10.0\GeVc$.

Calculations from theoretical models at midrapidity ($\abs{y}<1$) are also presented. 
These models use different assumptions of the QGP properties, for example in the 
thermal evolution of the collision system and in the initial-state conditions before the formation of the QGP. 
In addition, different mechanisms are considered regarding the interaction of heavy quarks with the medium 
and for the hadronization process. Results from the models LBT~\cite{Cao:2016gvr}, CUJET 3.0~\cite{Xu:2015bbz}, 
and SUBATECH~\cite{Nahrgang:2014vza} include collisional and radiative energy losses, 
while those from the models TAMU~\cite{He:2014cla}, PHSD~\cite{Liu:2012ax}, and TAMU SMCs~\cite{He:2019vgs} include only collisional energy loss. 
Initial-state fluctuations are included in the calculations by LBT, SUBATECH, and PHSD, and 
calculations for the $\vThree$ coefficient are only available from these three models. 
Coalescence mechanisms are also included in LBT, SUBATECH, TAMU, PHSD, and TAMU SMCs. 
While most models seem to capture the qualitative trend of the data (except for the $\vTwo$ description 
provided by TAMU in the 10--50\% centrality range), most of the models do not provide a quantitative description over the full range, 
except for TAMU SMCs. The TAMU SMCs version improves the TAMU model 
by implementing event-by-event space-momentum correlations (SMCs) between charm quarks and 
the high-flow partons in the QGP medium~\cite{He:2019vgs}. 
Since it does not include initial-state fluctuations, TAMU SMCs does not provide $\vTwo$ calculations for centrality values between 0--10\%. 
This puts more stringent constraints on the development of the collective flow for charm quarks in the QGP medium, 
giving further inputs for understanding heavy-quark interactions with the medium (for example, energy loss and coalescence mechanisms).

\begin{figure*}[thbp!]
  \centering
   \includegraphics[width=0.98\textwidth]{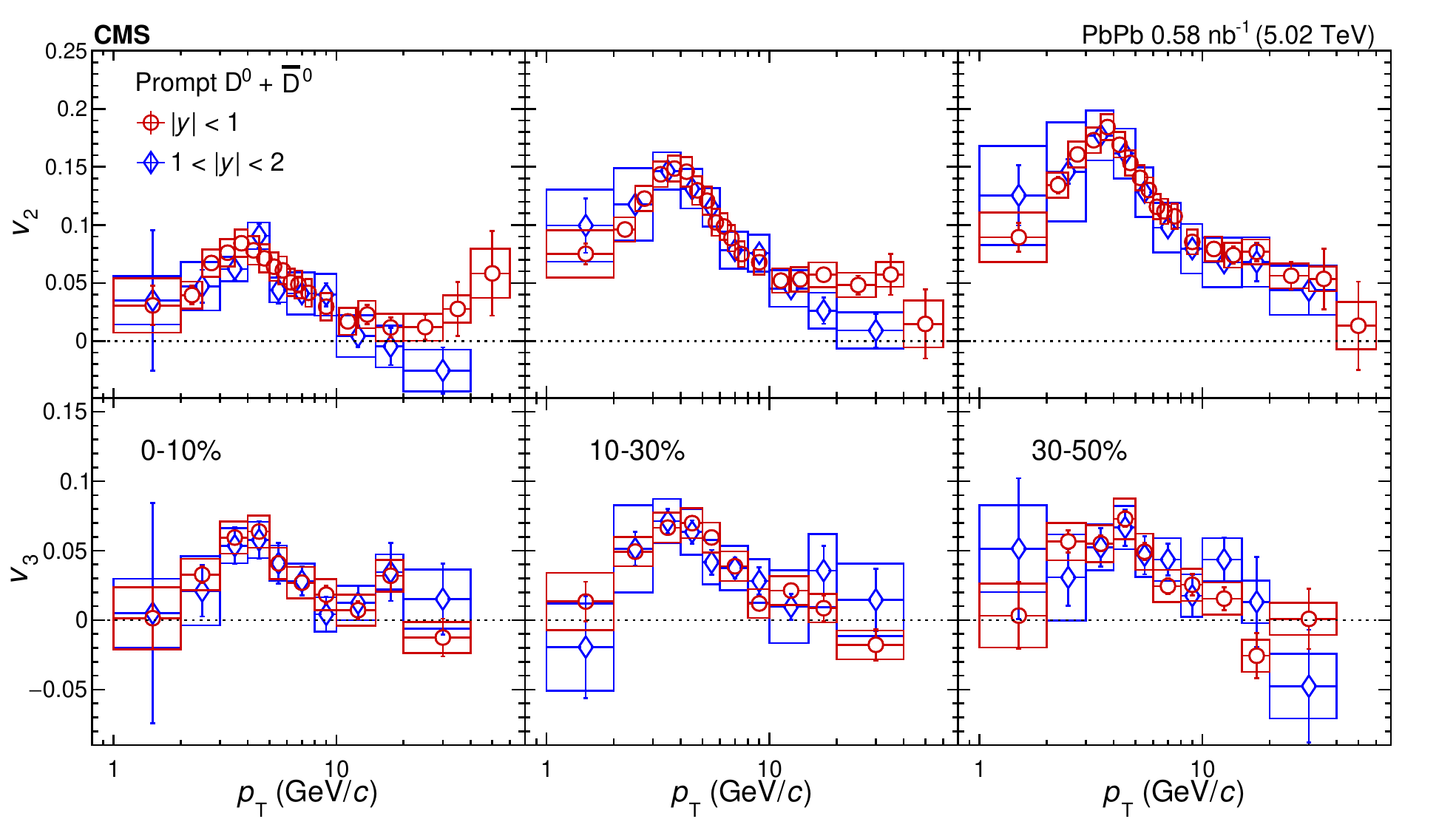}
  \caption{Prompt \PDz meson flow coefficients $\vTwo$ (upper) and $\vThree$ (lower) at midrapidity 
  ($\abs{y}<1$, red open circles) and forward rapidity ($1<\abs{y}<2$, blue open diamonds)
for the centrality classes 0--10\% (left), 10--30\% (middle), and 30--50\% (right). 
The vertical bars and open boxes represent the statistical and systematic uncertainties, respectively.
The horizontal bars represent the width of each \pt bin.
}
  \label{fig:results_vn_pTbins_forward}
\end{figure*}

Results for the rapidity dependence of heavy-flavor collective flow are presented for the first time 
for prompt \PDz meson $\vTwo$ and $\vThree$ as functions of \pt, both at midrapidity ($\abs{y}<1$) 
 and in the forward ($1<\abs{y}<2$) region, as shown in Fig.~\ref{fig:results_vn_pTbins_forward}.
No clear rapidity dependence is observed for both $\vTwo$ and $\vThree$ as functions of \pt.
This observation is similar to that for inclusive charged-hadron measurements~\cite{Sirunyan:2017igb}.

\begin{figure*}[!thb]p
  \centering
   \includegraphics[width=\textwidth]{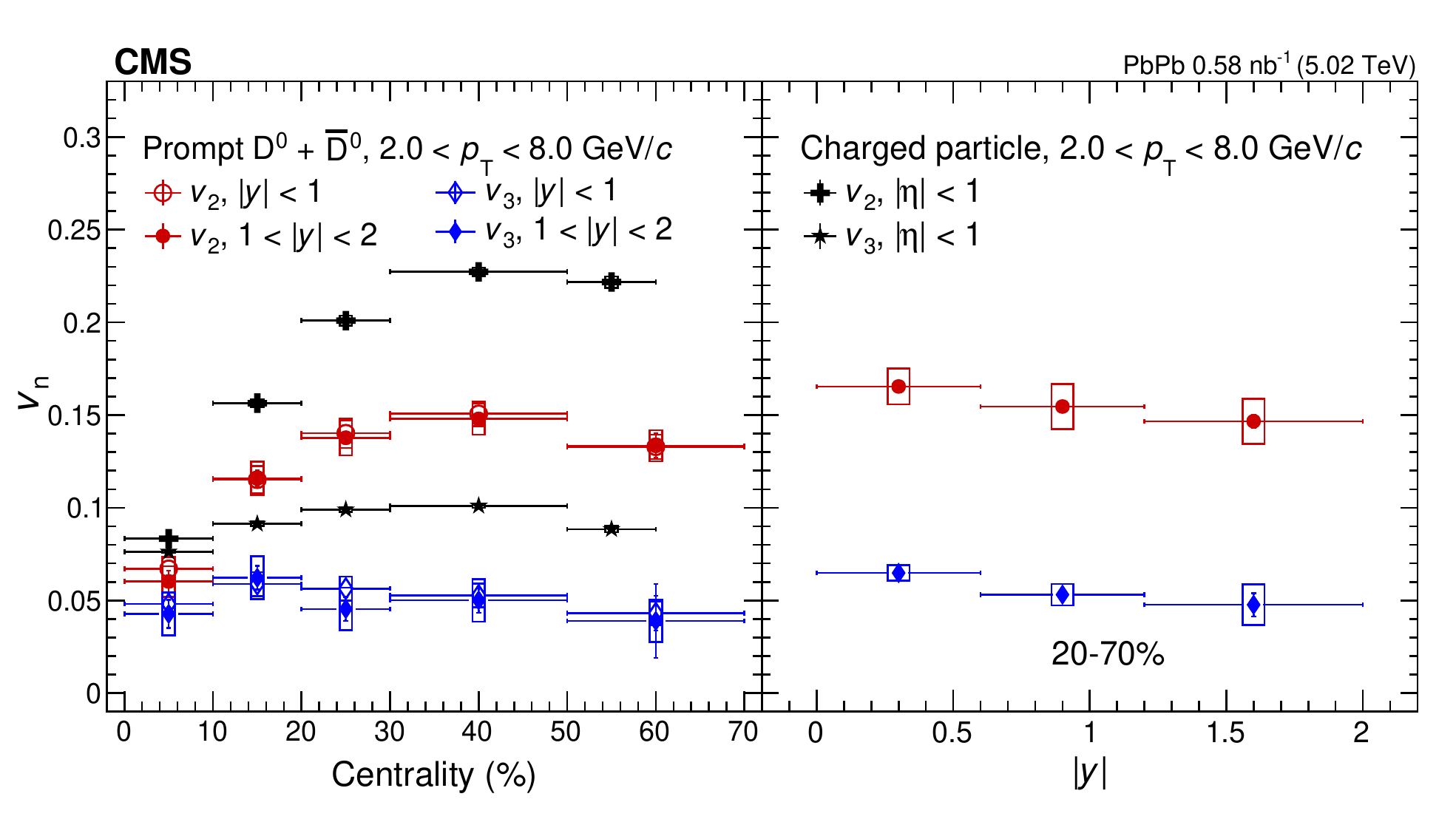}
  \caption{Prompt \PDz meson $\vTwo$ and $\vThree$ as functions of centrality, 
  for $2.0<\pt<8.0\GeVc$ and for rapidity ranges $\abs{y}<1$ and $1<\abs{y}<2$. The results are compared
with charged particle $\vTwo$ and $\vThree$ in the same \pt range and with $\abs{\eta}<1$ (left).  
Prompt \PDz  $\vTwo$ and $\vThree$ as functions of rapidity, for $2.0<\pt<8.0\GeVc$ and for centrality 20--70\% (right). 
The vertical bars represent statistical uncertainties and open boxes represent systematic uncertainties.
The horizontal bars represent the width of each bin.}
  \label{fig:results_vn_centbins}
\end{figure*}

In Fig.~\ref{fig:results_vn_centbins} (left), results for prompt \PDz mesons $\vTwo$ and $\vThree$,
averaged over $2.0<\pt<8.0\GeVc$, for $\abs{y}<1$ and $1<\abs{y}<2$, are presented as a function of
collision centrality. This \pt range is chosen in order to cover the widest possible \pt range, 
while maximizing the \PDz meson signal yield significance. 
These \pt- and rapidity-integrated results include an additional centrality bin (50--70\%), 
which has an insufficient number of events for the full differential analysis.
For both mid- and forward-rapidity regions, the $\vTwo$ results show
a clear increase from the most central to mid-central events, and then a declining trend 
toward the most peripheral events. 
This trend is similar to that observed for inclusive charged particles (also shown in Fig.~\ref{fig:results_vn_centbins}), 
and can be understood in terms of collision geometry and viscosity effects. In particular, a faster increase of $\vTwo$ 
is observed from central to peripheral collisions for charged particles compared to prompt \PDz mesons. 
This feature was also observed when comparing $\vTwo$ of low-\pt\ \JPsi with charged pions~\cite{Acharya:2020jil}, 
where it is claimed that this could be understood in terms of two phenomena: one, associated with transport models predicting 
an increasing fraction of regenerated \JPsi at low-\pt, when going from peripheral 
to central collisions; the other, not related to regeneration, is associated with a possible partial or 
later thermalization of charm quarks compared to light quarks~\cite{Acharya:2020jil}. The $\vThree$ shows
no centrality dependence, which is also consistent with expectations from collision geometry fluctuations~\cite{Chatrchyan:2013kba}.

\begin{figure}[!thbp]
  \centering
   \includegraphics[width=\cmsFigWidth]{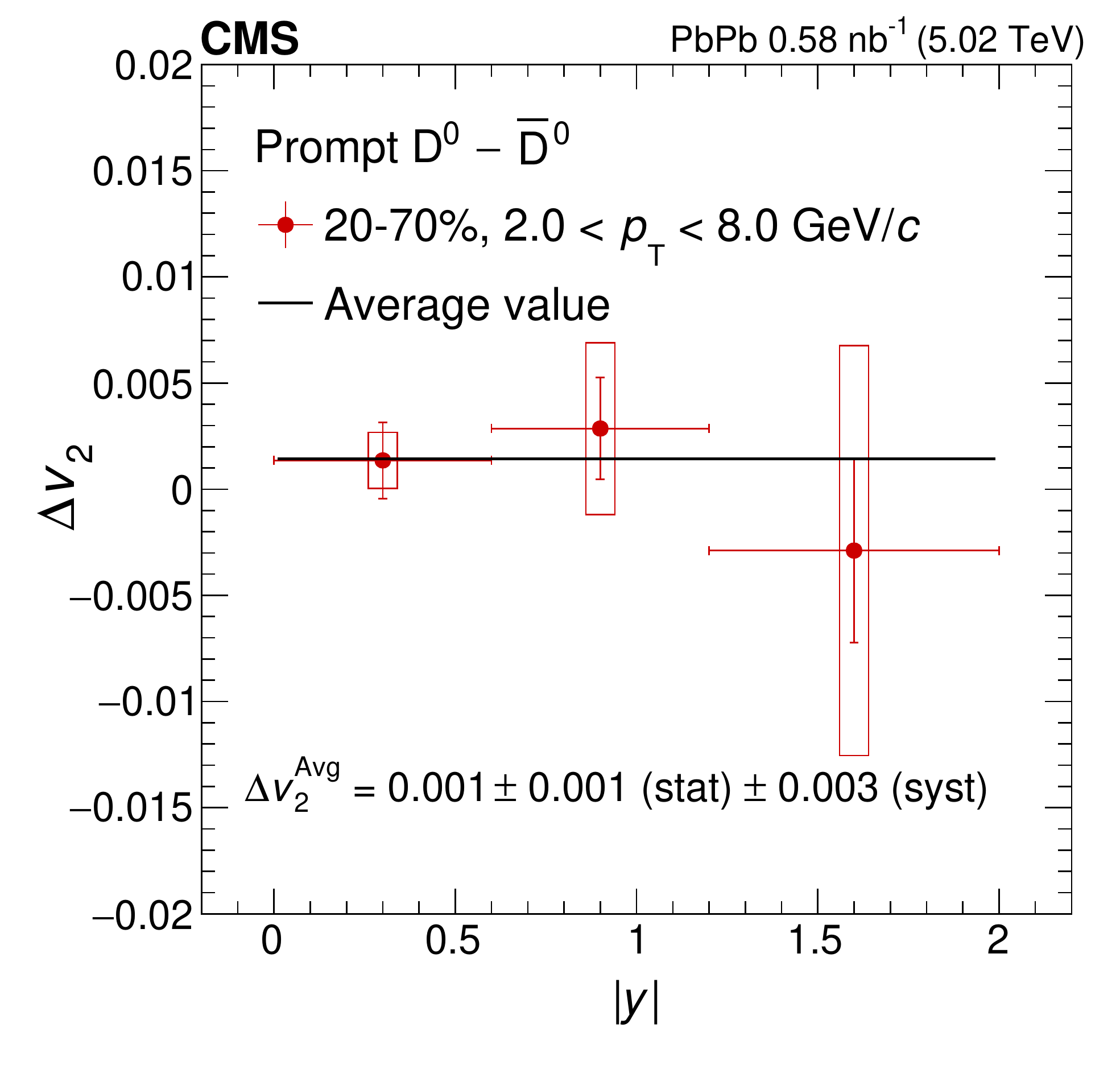} 
  \caption{Prompt \PDz meson $\Delta \vTwo$ as a 
  function of rapidity, for $2.0<\pt<8.0\GeVc$ and centrality 20--70\%.
The vertical bars represent statistical uncertainties and open boxes represent systematic uncertainties.  
The horizontal bars represent the width of each bin.
  }
  \label{fig:results_dvn_rapidity}
\end{figure}

Figure~\ref{fig:results_vn_centbins} (right) presents results for the rapidity dependence of 
prompt \PDz meson $\vTwo$ and $\vThree$, for centrality 20--70\%, averaged over
$2.0<\pt<8.0\GeVc$. A weak rapidity dependence of $\vTwo$ and $\vThree$ is observed in the data.

Finally, to search for effects of strong EM fields, the difference $\Delta \vTwo$ between the $\vTwo$ values of \PDz and \PADz mesons 
is measured. These results are presented in Fig.~\ref{fig:results_dvn_rapidity}, as a function of rapidity, averaged over $2.0<\pt<8.0\GeVc$ 
and for centrality 20--70\%. For all rapidity bins, the $\Delta \vTwo$ values are compatible with zero. 
The average over the full rapidity region is $\left\langle\Delta \vTwo \right\rangle = 0.001\pm0.001\stat\pm0.003\syst$. 
In Ref.~\cite{Gursoy:2018yai}, the predicted $\vTwo$ splitting for inclusive charged particles due to
electric fields is ${\sim}0.001$ at the LHC energies. While quantitative predictions for $\vTwo$ splitting of \PDz mesons are not yet available, 
they are expected to be much larger than those for inclusive charged particles. 
In the case of $\Delta v_1$, the ALICE collaboration reported results 
about three orders of magnitude larger than measurements for charged hadrons~\cite{Acharya:2019ijj}, 
although the uncertainties prevent a clear conclusion. 
The main reason is that heavy-flavor quarks are usually produced much earlier than light-flavor quarks, 
the former being predominantly produced soon after the collision takes place, 
when the EM field strength is several orders of magnitude stronger~\cite{Das:2016cwd}. 
The results presented here pose constraints on possible EM effects on charm quarks.

\section{Summary}
\label{summary}

Measurements of the elliptic ($\vTwo$) and triangular ($\vThree$) flow coefficients of prompt \PDz mesons
are presented as functions of transverse momentum (\pt), rapidity, and collision centrality,
in \PbPb collisions at $\sqrtsNN=5.02\TeV$. The results improve previously
published CMS data by extending the \pt and rapidity coverage and by providing
more differential information in \pt, rapidity, and centrality.
A clear centrality dependence of prompt \PDz meson $\vTwo$ is observed, while $\vThree$
is largely centrality independent. These trends are consistent with the expectation that
$\vTwo$ and $\vThree$ are driven by initial-state geometry. 
A weak rapidity dependence of prompt \PDz meson $\vTwo$ and $\vThree$ is observed. 
When comparing various theoretical calculations to the data at midrapidity,
no model is able to describe the data over the full centrality and \pt ranges. 

Motivated by the search for evidence of the strong electric field expected in \PbPb collisions, a 
first measurement of the $\vTwo$ flow coefficient difference ($\Delta \vTwo$) between \PDz and \PADz mesons
 as a function of rapidity is presented.  The rapidity-averaged $\vTwo$ difference is measured to be
$\langle\Delta \vTwo \rangle = 0.001\pm0.001\stat\pm0.003\syst$. 
This indicates that there is no evidence that charm hadron collective 
flow is affected by the strong Coulomb field created in ultrarelativistic heavy ion collisions. 
Future comparisons of theoretical models with these results 
may provide constraints on the electric conductivity of the quark-gluon plasma.

\begin{acknowledgments}
  We congratulate our colleagues in the CERN accelerator departments for the excellent performance of the LHC and thank the technical and administrative staffs at CERN and at other CMS institutes for their contributions to the success of the CMS effort. In addition, we gratefully acknowledge the computing centers and personnel of the Worldwide LHC Computing Grid for delivering so effectively the computing infrastructure essential to our analyses. Finally, we acknowledge the enduring support for the construction and operation of the LHC and the CMS detector provided by the following funding agencies: BMBWF and FWF (Austria); FNRS and FWO (Belgium); CNPq, CAPES, FAPERJ, FAPERGS, and FAPESP (Brazil); MES (Bulgaria); CERN; CAS, MoST, and NSFC (China); COLCIENCIAS (Colombia); MSES and CSF (Croatia); RIF (Cyprus); SENESCYT (Ecuador); MoER, ERC IUT, PUT and ERDF (Estonia); Academy of Finland, MEC, and HIP (Finland); CEA and CNRS/IN2P3 (France); BMBF, DFG, and HGF (Germany); GSRT (Greece); NKFIA (Hungary); DAE and DST (India); IPM (Iran); SFI (Ireland); INFN (Italy); MSIP and NRF (Republic of Korea); MES (Latvia); LAS (Lithuania); MOE and UM (Malaysia); BUAP, CINVESTAV, CONACYT, LNS, SEP, and UASLP-FAI (Mexico); MOS (Montenegro); MBIE (New Zealand); PAEC (Pakistan); MSHE and NSC (Poland); FCT (Portugal); JINR (Dubna); MON, RosAtom, RAS, RFBR, and NRC KI (Russia); MESTD (Serbia); SEIDI, CPAN, PCTI, and FEDER (Spain); MOSTR (Sri Lanka); Swiss Funding Agencies (Switzerland); MST (Taipei); ThEPCenter, IPST, STAR, and NSTDA (Thailand); TUBITAK and TAEK (Turkey); NASU (Ukraine); STFC (United Kingdom); DOE and NSF (USA).
    
  \hyphenation{Rachada-pisek} Individuals have received support from the Marie-Curie program and the European Research Council and Horizon 2020 Grant, contract Nos.\ 675440, 752730, and 765710 (European Union); the Leventis Foundation; the A.P.\ Sloan Foundation; the Alexander von Humboldt Foundation; the Belgian Federal Science Policy Office; the Fonds pour la Formation \`a la Recherche dans l'Industrie et dans l'Agriculture (FRIA-Belgium); the Agentschap voor Innovatie door Wetenschap en Technologie (IWT-Belgium); the F.R.S.-FNRS and FWO (Belgium) under the ``Excellence of Science -- EOS" -- be.h project n.\ 30820817; the Beijing Municipal Science \& Technology Commission, No. Z191100007219010; the Ministry of Education, Youth and Sports (MEYS) of the Czech Republic; the Deutsche Forschungsgemeinschaft (DFG) under Germany's Excellence Strategy -- EXC 2121 ``Quantum Universe" -- 390833306; the Lend\"ulet (``Momentum") Program and the J\'anos Bolyai Research Scholarship of the Hungarian Academy of Sciences, the New National Excellence Program \'UNKP, the NKFIA research grants 123842, 123959, 124845, 124850, 125105, 128713, 128786, and 129058 (Hungary); the Council of Science and Industrial Research, India; the HOMING PLUS program of the Foundation for Polish Science, cofinanced from European Union, Regional Development Fund, the Mobility Plus program of the Ministry of Science and Higher Education, the National Science Center (Poland), contracts Harmonia 2014/14/M/ST2/00428, Opus 2014/13/B/ST2/02543, 2014/15/B/ST2/03998, and 2015/19/B/ST2/02861, Sonata-bis 2012/07/E/ST2/01406; the National Priorities Research Program by Qatar National Research Fund; the Ministry of Science and Higher Education, project no. 02.a03.21.0005 (Russia); the Tomsk Polytechnic University Competitiveness Enhancement Program; the Programa Estatal de Fomento de la Investigaci{\'o}n Cient{\'i}fica y T{\'e}cnica de Excelencia Mar\'{\i}a de Maeztu, grant MDM-2015-0509 and the Programa Severo Ochoa del Principado de Asturias; the Thalis and Aristeia programs cofinanced by EU-ESF and the Greek NSRF; the Rachadapisek Sompot Fund for Postdoctoral Fellowship, Chulalongkorn University and the Chulalongkorn Academic into Its 2nd Century Project Advancement Project (Thailand); the Kavli Foundation; the Nvidia Corporation; the SuperMicro Corporation; the Welch Foundation, contract C-1845; and the Weston Havens Foundation (USA).
\end{acknowledgments}

\bibliography{auto_generated}
\cleardoublepage \appendix\section{The CMS Collaboration \label{app:collab}}\begin{sloppypar}\hyphenpenalty=5000\widowpenalty=500\clubpenalty=5000\vskip\cmsinstskip
\textbf{Yerevan Physics Institute, Yerevan, Armenia}\\*[0pt]
A.M.~Sirunyan$^{\textrm{\dag}}$, A.~Tumasyan
\vskip\cmsinstskip
\textbf{Institut f\"{u}r Hochenergiephysik, Wien, Austria}\\*[0pt]
W.~Adam, F.~Ambrogi, T.~Bergauer, M.~Dragicevic, J.~Er\"{o}, A.~Escalante~Del~Valle, R.~Fr\"{u}hwirth\cmsAuthorMark{1}, M.~Jeitler\cmsAuthorMark{1}, N.~Krammer, L.~Lechner, D.~Liko, T.~Madlener, I.~Mikulec, N.~Rad, J.~Schieck\cmsAuthorMark{1}, R.~Sch\"{o}fbeck, M.~Spanring, S.~Templ, W.~Waltenberger, C.-E.~Wulz\cmsAuthorMark{1}, M.~Zarucki
% ...
\vskip\cmsinstskip
\dag: Deceased\\
1:  Also at Vienna University of Technology, Vienna, Austria\\
% ...
\end{sloppypar}
\end{document}